\documentclass[]{spie}  

\usepackage{amsmath,amsfonts,amssymb}
\usepackage{graphicx}

\usepackage{bm}
\usepackage{pgfplots}
\usepackage{pgfplotstable}
\usepackage{tikz}
\usetikzlibrary{arrows,arrows.meta}
\usepackage{ifthen}
\usepackage[product-units = power,range-units = single]{siunitx}
\usepackage{subfig}
\pgfplotsset{compat=1.14}

\title{
Feasibility of a markerless tracking system based on optical coherence tomography
}

\author[a]{Matthias Schl\"{u}ter}
\author[a]{Christoph Otte}
\author[a]{Thore Saathoff}
\author[a]{Nils Gessert}
\author[a]{Alexander Schlaefer}
\affil[a]{Institute of Medical Technology, Hamburg University of Technology, Hamburg, Germany.}

\usepackage{color}

\makeatletter
\renewcommand{\fps@figure}{!tb}
\renewcommand{\fps@table}{!tb}
\makeatother
\begin{document}

\maketitle

\begin{abstract}

Clinical tracking systems are popular but typically require specific tracking markers. During the last years, scanning speed of optical coherence tomography (OCT) has increased to A-scan rates above \SI{1}{\mega\hertz} allowing to acquire volume scans of moving objects. Therefore, we propose a markerless tracking system based on OCT to obtain small volumetric images including information of sub-surface structures at high spatio-temporal resolution. In contrast to conventional vision based approaches, this allows identifying natural landmarks even for smooth and homogeneous surfaces. We describe the optomechanical setup and process flow to evaluate OCT volumes for translations and accordingly adjust the position of the field-of-view to follow moving samples. While our current setup is still preliminary, we demonstrate tracking of motion transversal to the OCT beam of up to \SI{20}{\milli\meter\per\second} with errors around \SI{0.2}{\milli\meter} and even better for some scenarios. Tracking is evaluated on a clearly structured and on a homogeneous phantom as well as on actual tissue samples. The results show that OCT is promising for fast and precise tracking of smooth, monochromatic objects in medical scenarios. 

\end{abstract}



\section{Introduction}

Tracking an object-of-interest and reconstructing its trajectory of motion is a particularly interesting problem in navigation of medical procedures. Examples include localization of target structures \cite{Wein2015}, guidance of surgical interventions \cite{rassweiler2014surgical}, or motion compensation \cite{ernst13}. Several marker-based approaches for tracking systems exist, including cameras using infra-red light and active or passive markers and tracking by electromagnetic fields using measuring coils. However, the former requires a steady line-of-sight and the latter is sensitive to field distortions \cite{franz14}. Additionally, it is necessary to attach a marker to the target in both cases. Other approaches use natural landmarks instead.  For example, systems employing structured light or time-of-flight measurements generate sequences of point clouds which can be registered \cite{placht12,li16}. Markerless approaches require a certain amount of inhomogenity in the images, typically gradients, shapes, or depth variations \cite{bouget2017vision}. This is not generally the case in medical applications,  e.g., smooth surfaces like liver and forehead do not exhibit clear natural landmarks if only a small part is accessible or high accuracy is required.

We consider optical coherence tomography (OCT) as an alternative optical imaging approach to obtain 3D information for tracking. OCT is an interferometric image modality with an imaging depth of approximately \SI{1}{\milli\meter} in scattering tissue and an axial resolution of typically \SIrange{5}{15}{\micro\meter}. Hence, OCT can resolve sub-surface microstructures for many biological tissues. Current systems allow for A-scan acquisition rates exceeding \SI{1.5}{\mega\hertz}, thus allowing for 4D real-time imaging \cite{Wang2016}. Previously, different motion compensation strategies based on OCT image volumes have been studied, e.g., to detect and compensate for small motion during ophthalmic interventions \cite{Kocaoglu:14,carrasco2015pupil} or laser cochleostomy \cite{zhang2014optical}. In contrast to conventional tracking systems, the field of view (FOV) in OCT is typically only in the range of a few millimeters. While OCT has also been used to scan larger ranges at high resolution \cite{finke2012automatic,rajput}, these approaches use robotic devices to mechanically move the scan head rather slowly. 

We propose using OCT directly for markerless tracking of a moving object-of-interest by exploiting arbitrary surface as well as sub-surface information. For this purpose, we describe an optomechanical setup to overcome the limited FOV size. Moreover, we demonstrate that homogeneous surfaces can be tracked using the proposed setup. We focus on evaluation of motion transversal to the OCT beam direction, because motion along beam direction leads to easily detectable shifts in the A-scans.

\section{Material and Methods}

\subsection{Tracking System Design}
Our tracking system is based on an OCT device providing volumetric images with high spatial and temporal resolution. A second scanning stage allows to move the FOV in space without moving the actual scan head. We use image processing to extract the motion of the sample and use this information to create a control loop which repositions the FOV such that a selected part of an object-of-interest is kept within our volumetric images while simultaneously recording the FOV position. 


We use a commercially available swept-source OCT system (OMES, OptoRes, Germany) with \SI{1.59}{\mega\hertz} A-scan rate and axial resolution of \SI{15}{\micro\meter} in air. The scan head in use allows for B-scan rates of about \SI{30}{\kilo\hertz}. Choosing a small sampling pattern of $32 \times 32$ A-scans for each C-scan, the setup allows acquisition of 833 volumes per second. We use sparse sampling in a lateral FOV of size \SI{3x3}{\milli\meter} to obtain information from larger structures as well. In axial direction, we have $476$ pixels per A-scan, covering a range of about \SI{3.5}{\milli\meter}. Thus, our FOV is \SI{3 x 3 x 3.5}{\milli\meter} and is sampled on a grid of \SI[product-units = single]{32 x 32 x 476}{voxels}.

\begin{figure}[!tb]
\centering
\begin{tikzpicture} \footnotesize
\node at (0,0) {\includegraphics[width=0.55\linewidth]{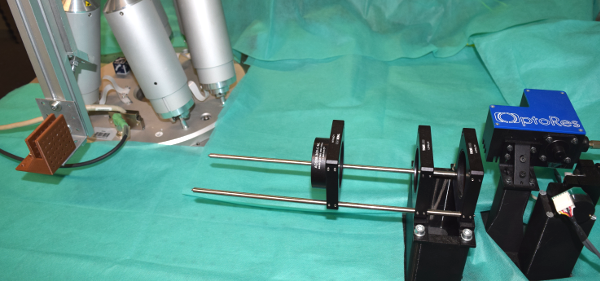}};
\node[draw=black,fill=white,circle,inner sep=1.5pt,minimum size=12] at (3.5,0.85) {a};
\node[draw=black,fill=white,circle,inner sep=1.5pt,minimum size=12] at (3.75,-1.4) {b};
\node[draw=black,fill=white,circle,inner sep=1.5pt,minimum size=12] at (0.5,-1.25) {c};
\node[draw=black,fill=white,circle,inner sep=1.5pt,minimum size=12] at (-3.75,-0.75) {d};
\node[draw=black,fill=white,circle,inner sep=1.5pt,minimum size=12] at (-0.75,1.25) {e};

\draw[-latex,white,thick] (-3.45,-0.05) node[above,yshift=0.3]{$x$} -- ++(1.1*0.3,1.1*0.4);
\draw[-latex,white,thick] (-3.45,-0.05) node[below,xshift=-1.6]{$y$} -- ++(-1.1*-0.25,-1.1*0.5);

\end{tikzpicture} \ 
\begin{tikzpicture}[scale=0.83]
\draw (0,0) -- (0,2) -- (3,2) node[midway,above] {scan head (a)} -- (3,0) -- (0,0);
\draw (3,0.5) rectangle (3.2,1.5);

\draw (6,2) -- ++(1.7,0) node[midway,above]{2nd galvo pair (b)} -- ++(0,-2) -- ++(-0.3,0) -- ++(0,1.7) -- ++(-1.4,0) -- (6,2);
\draw (6.5,1.7) rectangle (7,1.6);
\draw (7.4,0.75) rectangle (7.3,1.25);

\draw (6.75,1.6) -- ++(0,-0.1);
\draw (7.3,1) -- ++(-0.1,0);

\draw (6.75,1.5) -- ++(0.15,-0.1) -- ++(0,-0.7) -- ++(-0.15,-0.1) -- ++(-0.15,0.1) -- ++(0,0.7) -- ++(0.15,0.1);
\draw[fill=white] (7.2,1) -- ++(-0.1,0.15) -- ++(-0.7,0) -- ++(-0.1,-0.15) -- ++(0.1,-0.15) -- ++(0.7,0) -- ++(0.1,0.15);

\draw[->,very thick,>={Stealth[inset=0pt,length=8,angle'=28,round]}] (3.5,1.1) -- (6,1.1) node[midway,above]{C-scan beam};
\draw[<-,very thick,>={Stealth[inset=0pt,length=8,angle'=28,round]}] (3.5,0.9) -- (6,0.9);

\draw[->,very thick,>={Stealth[inset=0pt,length=8,angle'=28,round]}] (6.85,0.5) -- (6.85,-0.5);
\draw[<-,very thick,>={Stealth[inset=0pt,length=8,angle'=28,round]}] (6.65,0.5) -- (6.65,-0.5);

\node[draw,rounded corners=2,minimum width=2cm,minimum height=0.075cm] at (6.75,-0.75) {};
\node at (4.5,-0.75) {lens (c)};

\draw[->,very thick,>={Stealth[inset=0pt,length=8,angle'=28,round]}] (6.85,-1) -- (6.85,-2);
\draw[<-,very thick,>={Stealth[inset=0pt,length=8,angle'=28,round]}] (6.65,-1) -- (6.65,-2);

\draw (6.75,-2.5) ellipse (0.8 and 0.4) {};
\draw[very thin] (6.75,-2.5) ellipse (0.7 and 0.35) {};
\draw[very thin] (6.75,-2.5) ellipse (0.6 and 0.3) {};
\draw[ultra thin] (6.75,-2.5) ellipse (0.5 and 0.25) {};
\fill[white] (6.15,-2.75) rectangle (7.35,-2.28);
\fill[white] (6.75,-2.7) ellipse (0.6 and 0.17) {};
\fill[white] (6.13,-2.5) ellipse (0.1 and 0.19) {};
\fill[white] (7.38,-2.5) ellipse (0.1 and 0.19) {};
\node at (5,-2.5) {sample};
\end{tikzpicture}
\caption{
Experimental setup for lateral FOV positioning: scan head creating a C-scan pattern (a), second setup of galvo mirrors for lateral shifting (b), and an achromatic lens (c). The left image additionally shows the sample holder (d) attatched to a hexapod robot (e) to simulate lateral motion (white $xy$-frame).
}
\label{f:galvos}
\end{figure}
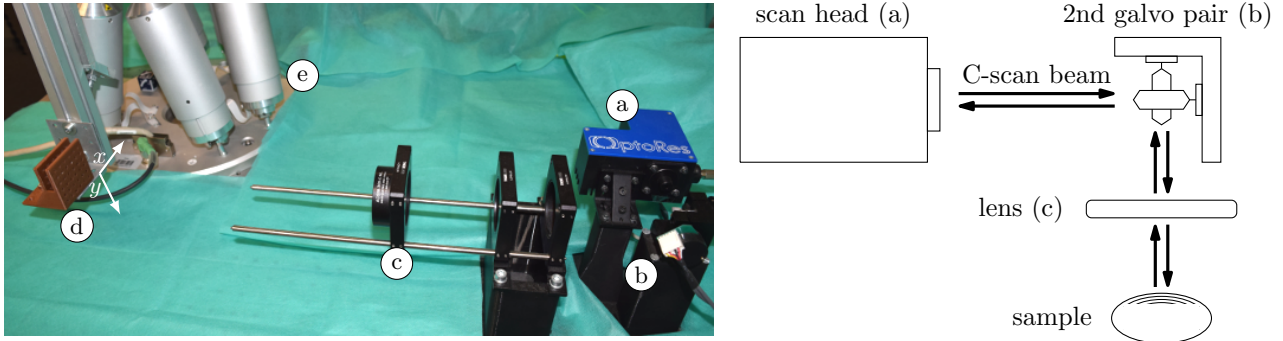

 
Positioning of the small FOV in a larger space is separated into axial and lateral translation. Axial position of the scan area can be changed by varying the pathlength in the reference arm. This is realized by translating a mirror with a stepper motor. For lateral repositioning, we use, as shown in Fig.~\ref{f:galvos}, a second galvo stage in front of the scan head and a focussing lens, whose diameter of \SI{6}{cm} limits the positioning range. Galvos and stepper are controlled by a microcontroller communicating with our measurement computer. A step of a galvo of the second stage corresponds to a shift of about \SI{30}{\micro\meter}. Although we do not explicitly evaluate axial motion in this work, variations along this axis are compensated, too.

To estimate the motion from two subsequent OCT volumes, their spatial relation needs to be determined. While for OCT several sophisticated feature extraction and matching methods have been proposed \cite{gan2014automated,niemeijer2009,Laves17}, computational effort is critical for our application. Therefore, we use the phase correlation method, which is a frequency domain approach to find the translation which maximizes cross-correlation between a signal $d(x,y,z)$ and a template $t(x,y,z)$.  
If we assume a pure translational motion, i.e.,
 $d(x,y,z) = t(x-\tau_x, y-\tau_y, z-\tau_z)$, then phase correlation calculates 
\begin{equation}
r(x,y,z) = \mathcal{F}^{-1} \left\{ \frac{\hat{d}(u,v,w) \cdot \hat{t}^\ast(u,v,w)}{|\hat{d}(u,v,w) \cdot \hat{t}^\ast(u,v,w)|} \right\},
\end{equation}
with hats denoting the Fourier transforms of $d$ and $t$ and $\mathcal{F}^{-1}$ being the inverse Fourier transform, and the resulting function $r(x,y,z)$ has a peak at the position corresponding to the translation $\begin{pmatrix}\tau_x, \tau_y, \tau_z \end{pmatrix}$ between the volumes. The involved operations can be implemented efficiently on a GPU.
The template in our case is an arbitrary initial OCT volume of a part of the object-of-interest. The determined translation leads to repositioning of the OCT's FOV in order to follow the object and keep it in focus. Hence, we need to convert the translation, which is in our implementation an integer number of voxels for each axis, into the corresponding number of motor steps. The steps are then  multiplied with a gain factor $g<1$ to smooth repositioning and especially avoid overshooting during tracking.

\subsection{Experimental Setup}
\begin{figure}[!tb] \centering
\subfloat[]{
\includegraphics[width=0.3\linewidth]{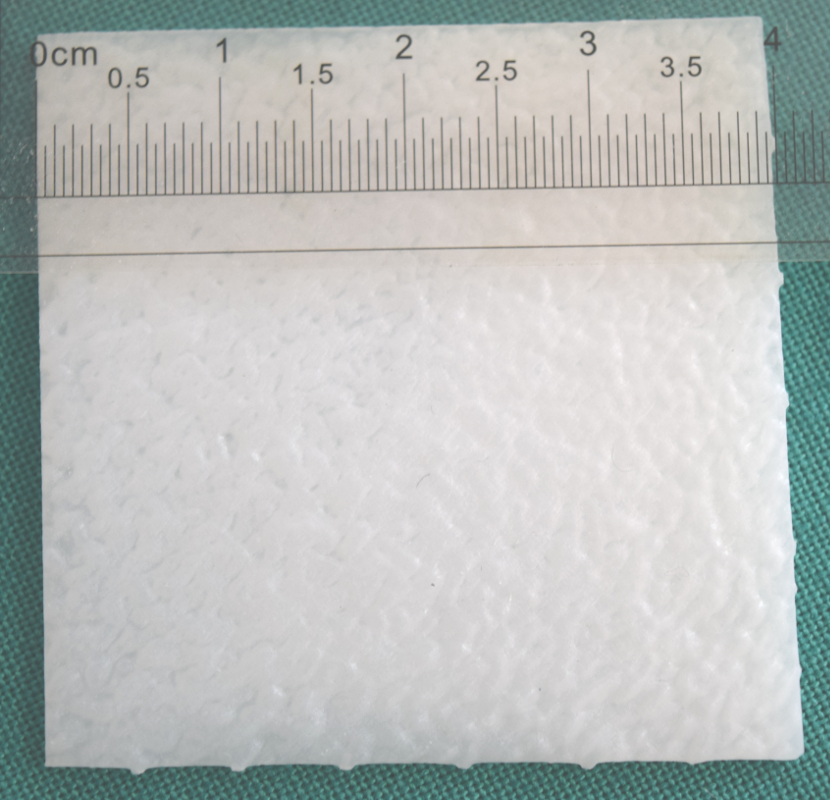} 
} \qquad
\subfloat[]{
\begin{tikzpicture}[scale=1.3] \footnotesize
        \begin{axis}[
            at={(0,0)},
            width=5cm,
            enlargelimits=false,
            axis on top,
            axis equal image,
            ticks=none
        ]
        \addplot graphics[xmin=0,xmax=40,ymin=0,ymax=40] {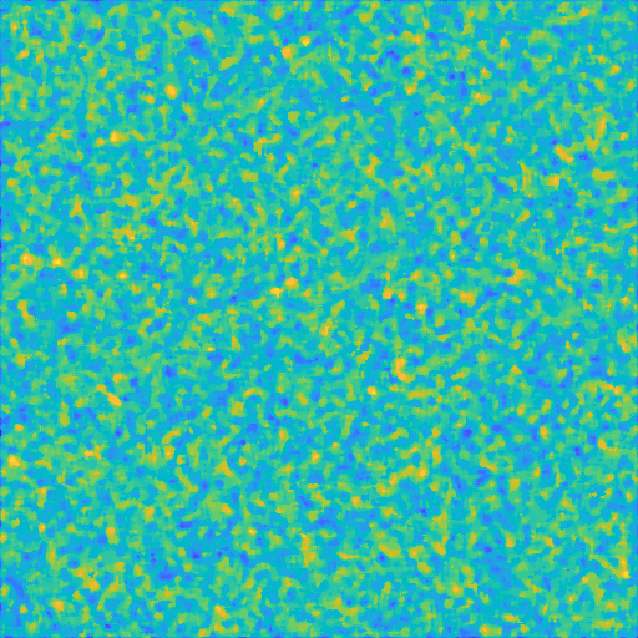};
        \end{axis}
        
        \begin{axis}[
            at={(0,50)},
            width=4.3cm,
            enlargelimits=false,
            axis on top,
            axis line style={draw=none},
            ytick=\empty,
            axis equal image,
            ticklabel shift={-6.5mm},
            xlabel shift={-7mm},
            xlabel=\si{\milli\meter},
            xlabel near ticks
        ]
        \addplot graphics[ymin=0,ymax=0.03,xmin=0,xmax=0.4531] {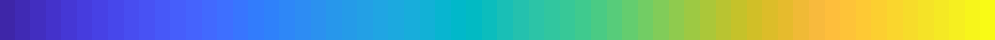};
        \end{axis}
\end{tikzpicture}
} \qquad
\subfloat[]{
\includegraphics[width=0.25\linewidth]{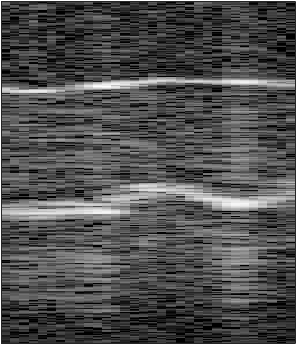}
}
\caption{3D-printed plate phantom made of resin as a picture (a), a relative depthmap of its fine structured surface in the underlying CAD model (b), and an exemplary B-scan (c)}
\label{f:plate}
\end{figure}

\begin{figure}[!tb] \centering
\subfloat[]{
\includegraphics[width=0.275\linewidth]{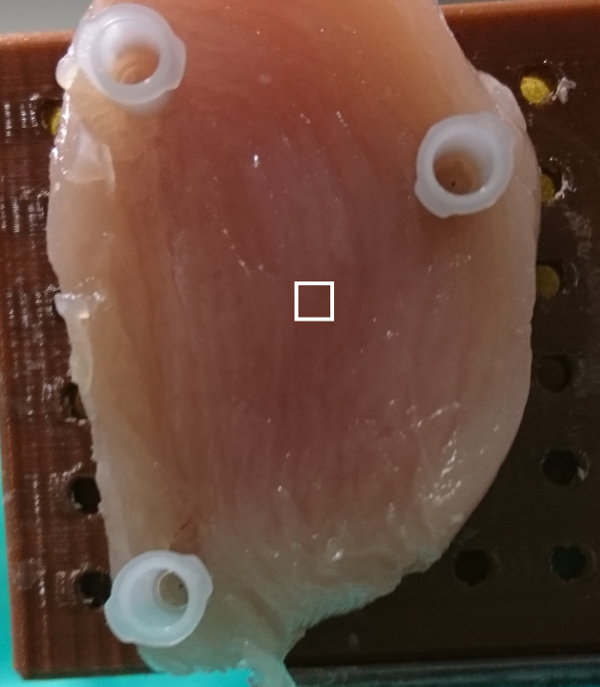}
} \qquad \subfloat[]{
\includegraphics[width=0.33\linewidth]{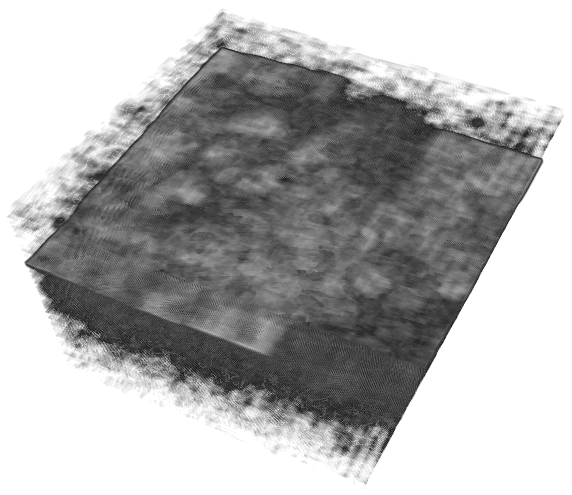}
}
\caption{Chicken breast sample pinned to the sample holder with a white rectangle illustrating the \SI{3x3}{\milli\meter} lateral FOV (a). Rendered volume scan of the sample, mainly showing its surface (b).}
\label{f:chicken}
\end{figure}

For evaluation, we move samples periodically with a hexapod robot (H-820, Physik Instrumente, Germany). The robot has a repeatability of less than \SI{20}{\micro\meter} for translational movements and a maximum velocity of \SI{20}{\milli\meter\per\second}. The baseline phantom is made of polyoxymethylene (POM) and consists of a plate with an extruding cube with a side length of \SI{1}{\milli\meter}. Its surface gives a strong OCT signal and motion of the cube should be easy to detect. As a second phantom, we use a 3D-printed plate as illustrated in Fig.~\ref{f:plate}. It is made of photopolymer resin and is designed with a fine, randomly structured surface on the top and bottom side. Its thickness is about \SI{1.4}{\milli\meter} such that reflections from both sides are visible in the OCT scan. Lastly, we use cut samples of chicken breast to evaluate tracking of actual tissue. The FOV is positioned at arbitrary central positions as sketched in Fig.~\ref{f:chicken}.

For the experiments, the hexapod moves the phantom and records its positions. The position of the galvos and the stepper motor are recorded as well. A simple calibration sequence is used to determine the spatial transformation between hexapod frame and galvo/motor steps. Thus, we are able to move the hexapod along the axes defined by the galvos/motor as well as to evaluate the tracking error in millimeter. Additionally, the timestamps of both systems are synchronized. 54 volumes per second are evaluated for repositioning of the FOV.

\section{Results and Discussion}
Subsequently, $x$- and $y$-axis refer to the axes defined by the second galvo pair (Fig.~\ref{f:galvos}). They are approximately parallel to the lateral axes of the OCT volumes. For all experiments, the hexapod moves the sample repeatedly back and forth over a distance of \SI{20}{\milli\meter} along these axes. The tracking is labeled as a failure, if the sample is lost and the FOV is moved completely away from the target. Typically this motion is random until the motor limits are reached and therefore easily detectable. 

\begin{figure}[!tb]
\centering
\subfloat[Cube, along $x$, one failure at \SI{16}{\milli\meter\per\second} and two at \SI{20}{\milli\meter\per\second}]{
\includegraphics[trim={20 0 35 22},clip,width=0.46\textwidth]{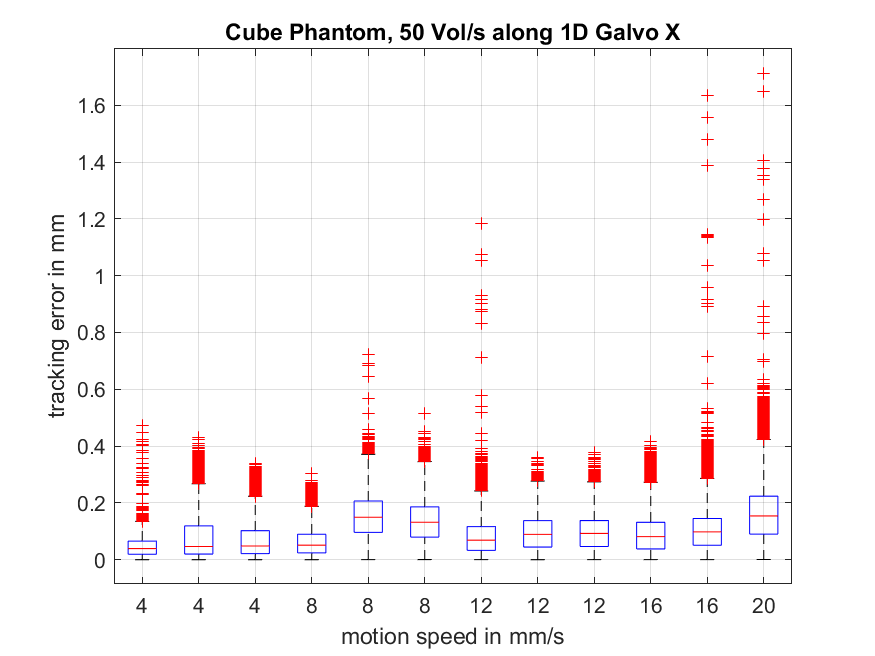}
} \qquad
\subfloat[Cube, along $xy$, one failure at \SI{20}{\milli\meter\per\second}]{
\includegraphics[trim={20 0 35 22},clip,width=0.46\textwidth]{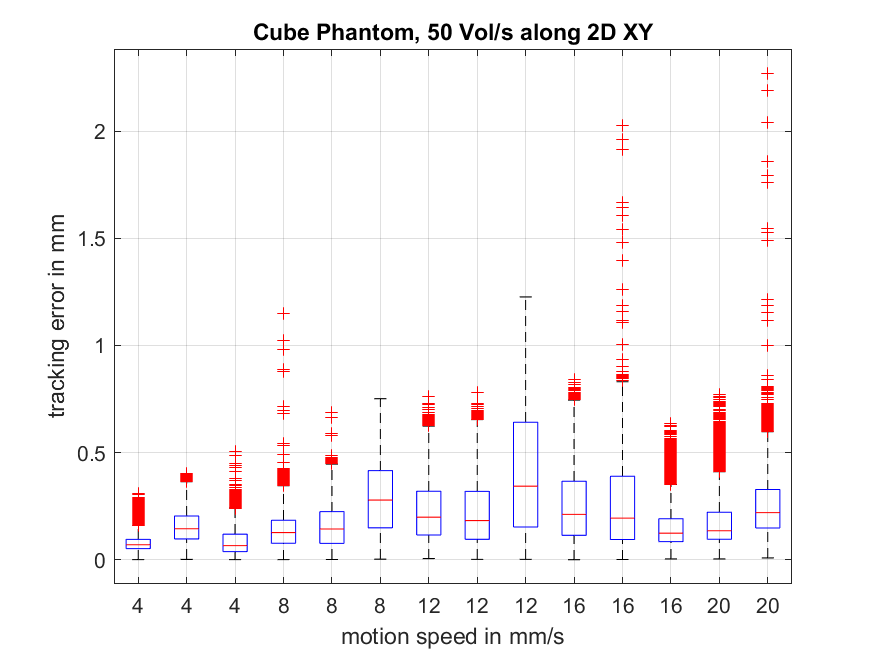}
}

\subfloat[Plate, along $x$, no failures]{
\includegraphics[trim={20 0 35 22},clip,width=0.46\textwidth]{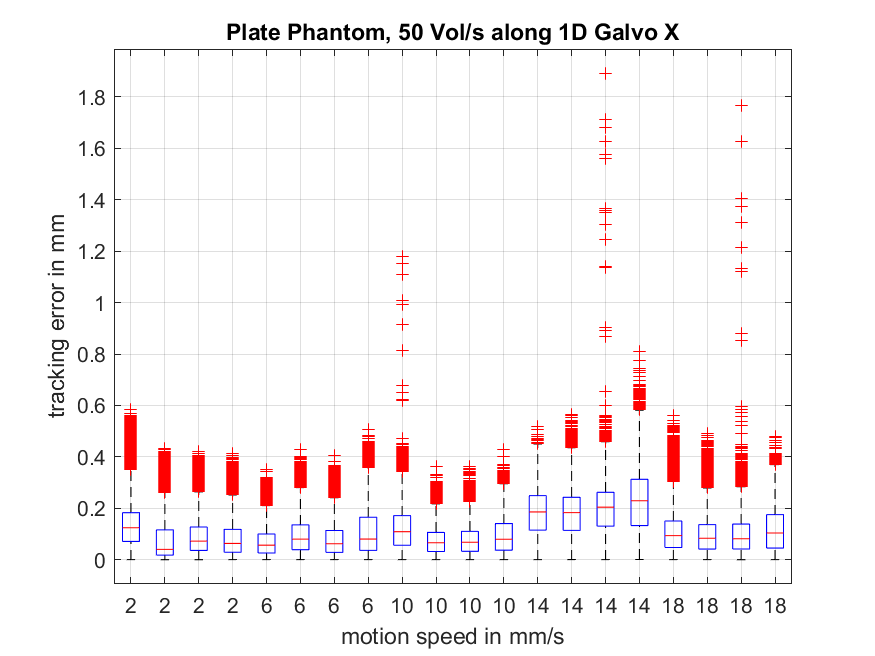}
} \qquad
\subfloat[Plate, along $xy$, two failures at \SI{18}{\milli\meter\per\second}]{
\includegraphics[trim={20 0 35 22},clip,width=0.46\textwidth]{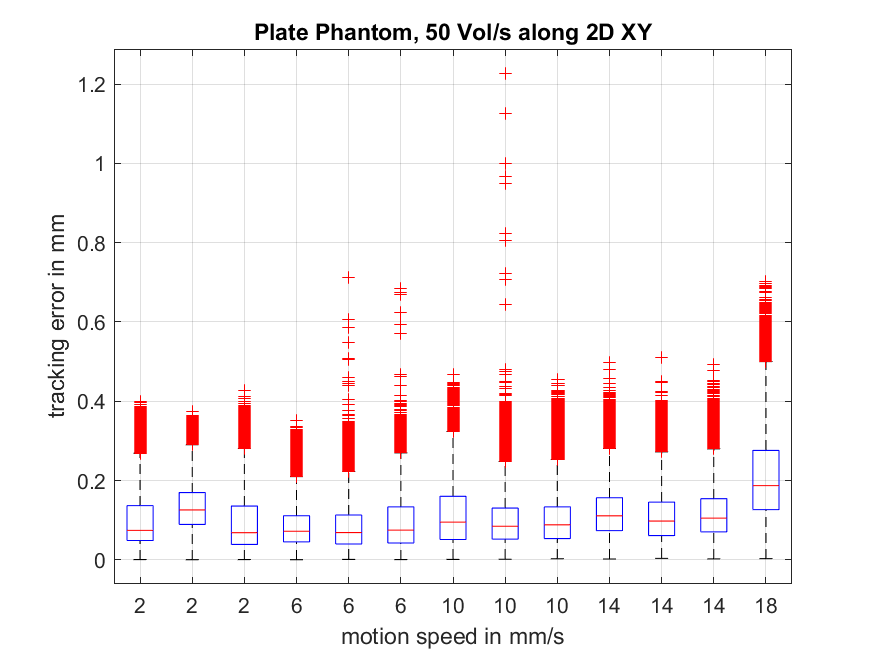}
}
\caption{Tracking errors based on Euclidean distances for the cube phantom along $x$- and $xy$-direction (a,b) and for the plate phantom along $x$ and $xy$ (c,d). Each box is based on at least \num{14758} positions logged during at least \SI{3}{\minute} of continuous tracking.}
\label{f:cubeplate}
\end{figure}

Fig.~\ref{f:cubeplate} shows box-and-whisker plots of the tracking errors for the cube and plate phantom with varying speed along $x$ and along $x$ and $y$ simultaneously. Here, errors are calculated as Euclidean distances between the tracked positions and the actual hexapod positions. Tracking was possible for up to \SI{20}{\milli\meter\per\second}.  For tracking of the cube phantom along $xy$, the errors are slightly higher than for the other three scenarios. While for both phantoms the median error is mostly below \SI{0.2}{\milli\meter}, some high outliers and tracking failures occured, however, and might require proper handling in an application.


%

\begin{figure}[!tb]\centering
\subfloat[]{
\includegraphics[width=0.5\textwidth]{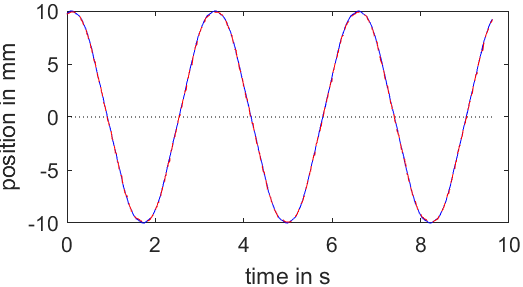}
}
\subfloat[]{
\begin{tikzpicture}
\node at (0,2.7) {\includegraphics[height=0.15\textwidth]{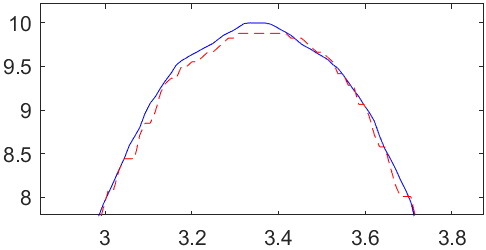}};
\node at (0.07,0) {\includegraphics[height=0.15\textwidth]{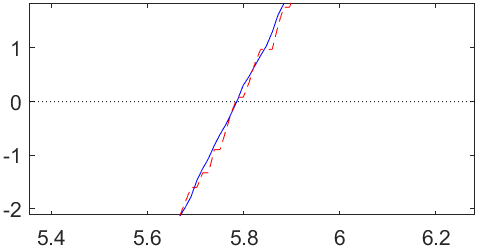}};
\end{tikzpicture}
}
\caption{An exemplary \SI{10}{\second} window of the tracked (red) and hexapod (blue) trajectory for the chicken breast sample along $x$ at \SI{18}{\milli\meter\per\second} (a). Additionally, zoomings of one of the turning points and a zero crossing are shown (b).}
\label{f:traj}
\end{figure}

For the chicken case, average root-mean-square (RMS) tracking errors are shown in Tab.~\ref{t:chicken_t} for motion along $x$ as well as $x$ and $y$ simultaneously. Each speed was tested on six different samples. From the 60 tests, tracking failed in 9 cases and in 4 tests an RMS above \SI{0.7}{\milli\meter} occured, maybe due to tracking of a similar region close-by instead of the actual template. The RMS for $xy$ motion is on average \SI{0.107}{\milli\meter} higher than for only $x$ motion. For the latter, Fig.~\ref{f:traj} shows an exemplary direct comparison between tracked positions and logged hexapod positions during a short time window.

\begin{table}[!tb] \centering 
\caption{Mean RMS tracking errors based on 6 chicken breast samples moved along $x$ and $xy$. Failing and outlier samples, i.e., $\text{RMS} > \SI{0.7}{\milli\meter}$, were removed and their number is given in brackets. For each measurement, \SIrange{14598}{18594}{positions} were acquired during \SIrange{2.93}{3.73}{\minute}. }
\label{t:chicken_t}
\begin{tabular}{c|c|c|}
 & \multicolumn{2}{c|}{$\text{RMS} \pm \sigma_\text{RMS}$  \si{\milli\meter} \ \ [\# removed]}  \\ 
  \si{\milli\meter\per\second} & $x$ & $xy$ \\  \hline
2 & $0.107 \pm 0.062$ \ \ [1]  & $0.203 \pm 0.078$ \ \ [0] \\ 
6 & $0.096 \pm 0.057$ \ \ [0]  & $0.197 \pm 0.090$ \ \ [1] \\ 
10 & $0.093 \pm 0.058$ \ \ [0]  & $0.304 \pm 0.092$ \ \ [2] \\ 
14 & $0.222 \pm 0.140$ \ \ [2]  & $0.227 \pm 0.132$ \ \ [2] \\ 
18 & $0.134 \pm 0.089$ \ \ [2]  & $0.257 \pm 0.159$ \ \ [3] 
\end{tabular}%
\end{table}


\section{Conclusion}
We showed that optical coherence tomography is a promising imaging modality for markerless tracking of tissue. In our phantom study, both an extruding cube and a smooth plate could be tracked with median errors in the order of \SI{0.1}{\milli\meter}, which is a typical target error for optical tracking systems. Additionally, by evaluating chicken breast samples we showed that the setup is also suitable for markerless tracking of real tissue. Except for the cube phantom, no larger-scale surface structure was present, which could have been exploited by conventional tracking systems.

In this work, we effectively used less than \SI{10}{\percent} of the OCT's maximum volume rate. Therefore, tracking of substantially faster motion than our reported \SI{20}{\milli\meter\per\second} should be feasible in the future, making OCT attractive for applications requiring both high spatial accuracy and temporal resolution and where using artificial markers is impractical or impossible.




\bibliography{trackingrefs}
\bibliographystyle{spiebib} 

\end{document}